\documentclass[preprint,prb,showkeys,showpacs,preprintnumbers,amsmath,amssymb]{revtex4}
\usepackage{graphicx}
\usepackage{dcolumn}
\usepackage{bm}
\usepackage{subfigure}
\begin{document}

\hyphenation{%
rea-sons
}
\title{Polarization, piezoelectric constants and elastic constants of ZnO, MgO and CdO}
\author{Priya Gopal and Nicola A. Spaldin \\ Materials Department, University of California,\\ 
Santa Barbara, CA 93106-5050. }

\begin{abstract}
We report first principles density functional calculations of the spontaneous polarizations, piezoelectric 
stress constants  and elastic constants for the II-VI wurtzite-structure oxides ZnO, MgO and CdO. 
Using our pseudopotential self-interaction corrected implementation of density functional theory, we obtain 
polarization values of -0.05, -0.17 and -0.10 C/m$^2$, and piezoelectric constants, $e_{33}$ ($e_{31}$) of 1.34
(-0.57), 2.26 (-0.38) and 1.67 (-0.48) C/m$^2$ for structurally relaxed ZnO, MgO 
and CdO respectively. These properties are consistently larger in magnitude than the corresponding GaN, AlN 
and InN analogues. Therefore we predict that larger internal fields will be attainable in ZnO-based polarization 
field effect transistors than in equivalent GaN-based devices. 
\end{abstract}

\maketitle

Over the past decade, nitride-based III-V semiconductors have emerged as leading contenders for
many technological applications, particularly in  optoelectronics (blue and green lasers and 
light emitting diodes ~\cite{Nakamura,Nakamura1}), and microelectronics (high electron mobility 
transistors~\cite{Mishra}). Their utility in optoelectronics stems in large part from
their wide range of direct band gaps - 0.7 eV in InN, 3.4 eV in GaN, and 6.3 eV in AlN - combined 
with their ability to form complete solid solutions, which provides the entire
spectrum of emission wavelengths from 1800 nm (infra-red) to 200 nm (ultra-violet).
In addition, the macroscopic polarization associated with the wurtzite crystal structure plays
an important role in determining the electrical and optical properties. This is particularly important 
in heterostructures where a difference in polarization between layers induces an electric field at the 
interface, which in turn is screened by the formation of a two-dimensional electron gas (2DEG). The electron
mobility in such polarization-induced 2DEGs can be much higher than that in traditional impurity-doped
systems, in which the electrons scatter from ionized impurities. Thus high speed transistors known as POLFETs 
(polarization field effect transistors) can be produced~\cite{Mishra}.
 
Presently, there is growing interest in ZnO-based materials as alternatives to the nitrides. 
Indeed many properties of ZnO are similar, if not superior to those of GaN. 
Perhaps most importantly, and in striking contrast to GaN, 
single crystals of ZnO can be grown readily~\cite{ZnO-growth}, and used as substrates for the growth of
thin film devices. This facilitates the production of much higher quality films using homoepitaxy, 
and circumvents the problems associated with dislocation formation from epitaxial mismatch that plagues 
GaN growth. In addition, the exciton binding energy of ZnO (60 meV) is twice 
as large as that of GaN, leading to stronger electron-hole recombination  and thus improved 
optical efficiency~\cite{Reynolds}. Finally, the band gap of 
ZnO (3.4 eV) is very close to that of GaN, and it can likewise be varied systematically by alloying 
with MgO~\cite{Ohtomo,Ohtomo2} or CdO\cite{zncdo}. However, since MgO and CdO do not 
occur in the wurtzite structure, values of the electronic and elastic properties, crucial for the
modeling of heterojunction device performance\cite{2DEG1}, cannot be obtained experimentally.

In this work we calculate the spontaneous polarizations, piezoelectric coefficients and elastic
constants of wurtzite-structure MgO, ZnO and CdO. We calculate the optimized structures of the three
materials using the standard local density approximation (LDA) to density functional theory as
implemented in the VASP package\cite{vasp1,vasp2}. We then use our recently-developed pseudopotential 
self-interaction corrected (pseudo-SIC) implementation~\cite{Alessio-Spaldin} of the density functional 
formalism to calculate the electronic properties, because the LDA fails to obtain a band gap for 
wurtzite-structure CdO, preventing the calculation of its polarization. The polarizations are obtained 
using the widely-used Berry phase method \cite{Vander-pol1,Vander-pol2}, and the piezoelectric coefficients 
and elastic constants are obtained from the calculated dependence of polarization and energy on the 
appropriate strain. Our polarizations are reported relative to the conventional reference 
structure of the zincblende structure with the same ratio of $c$ lattice constant to unit 
cell volume\cite{Bernardini}, and care is taken to avoid problems associated with the branch dependence 
of the polarization by extracting the so-called {\it proper} piezoelectric response for the $e_{31}$ 
coefficients\cite{Vanderbilt_ProperPE}. Our main result is that the 
polarization magnitudes and polarization gradients are all larger in the 
ZnO-based system than in the GaN-based system, suggesting ZnO as a promising candidate for 
the development of polarization field-effect transistors.

\begin{figure}[htbp]
\begin{center}
\includegraphics*[width=0.45\textwidth]{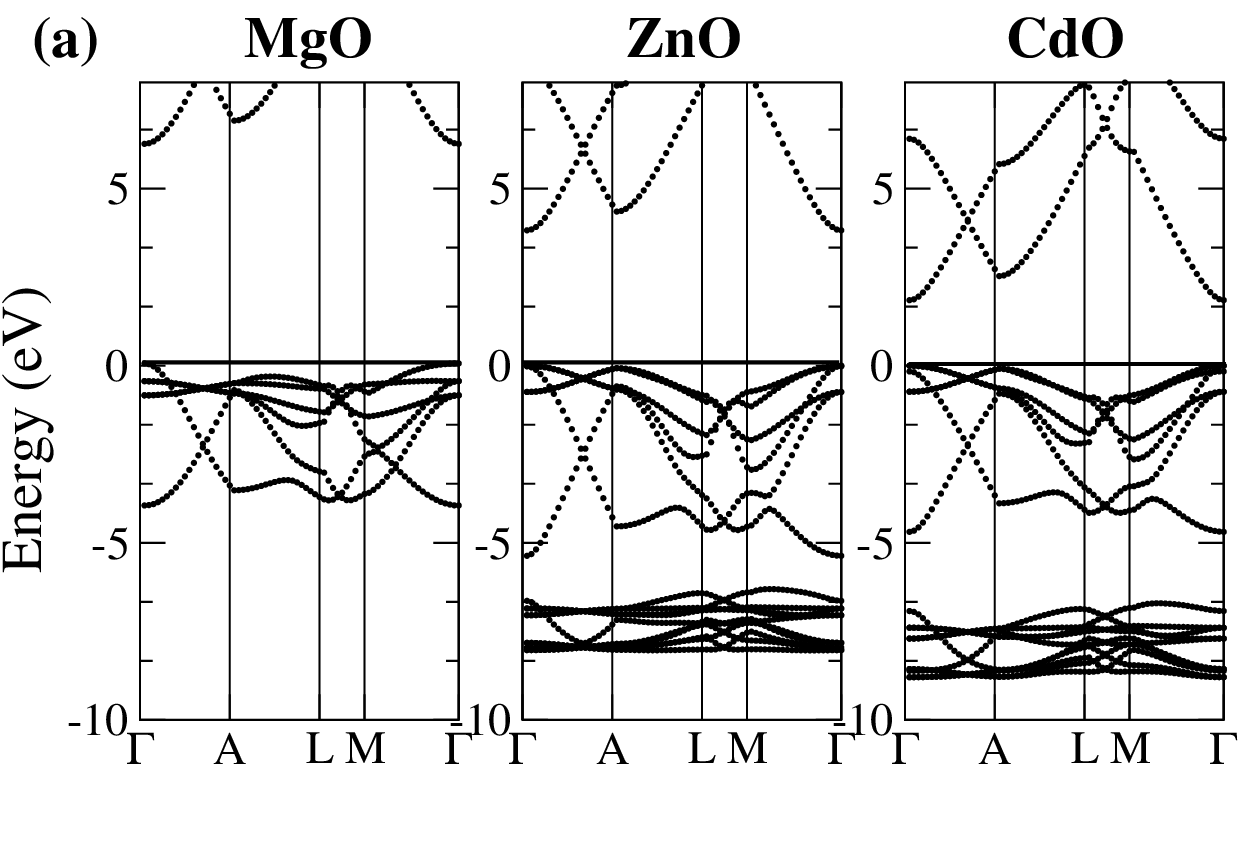} 
\includegraphics*[width=0.45\textwidth]{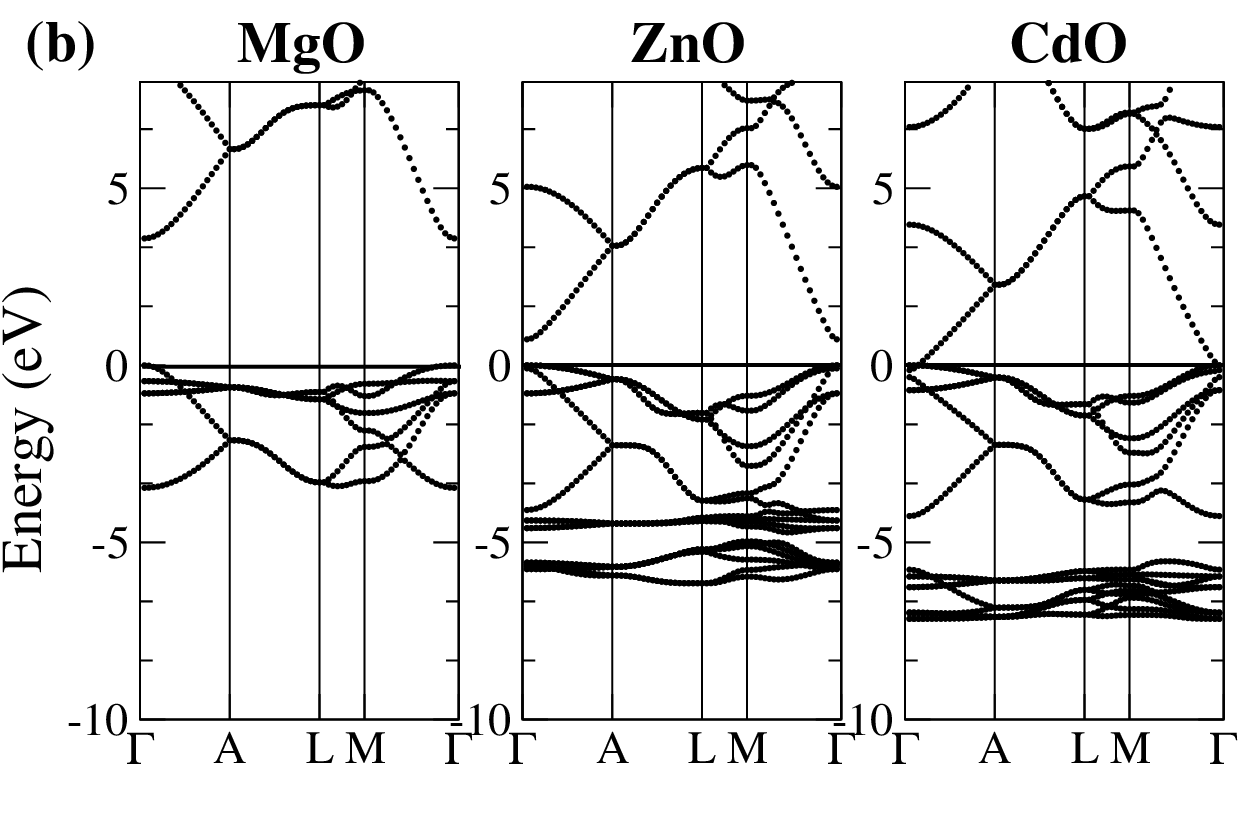}
\caption{Band structures of MgO, ZnO and CdO within (a) pseudo-SIC and (b) LDA approaches, at the 
calculated optimized LDA lattice parameters. Notice that the LDA severely underestimates the band gap, 
and results in a metallic ground state for CdO; this prevents the calculation of the polarization and 
piezoelectric coefficients.}
\label{lda}  
\end{center}
\end{figure}

Before reporting our polarization results, we compare our pseudo-SIC band structures with those 
that we obtain using the LDA (Figure \ref{lda}), at the calculated LDA lattice parameters. It is clear 
that our calculated pseudo-SIC band gap of ZnO (3.8 eV) is in good agreement with the experimental 
value (3.4 eV \cite{LB}), whereas the LDA shows the usual underestimation (0.78 eV). CdO and MgO 
occur naturally in the rock-salt structure with band gaps of $\sim$ 1 eV (indirect) or 2.4 eV (direct)
for CdO\cite{LB} and $\sim$ 7.7 eV for MgO\cite{LB}; although our calculated band gaps for the 
wurtzite structure should not match exactly, they are of a similar magnitude. 

When a wurtzite-structure semiconductor is constrained by epitaxial matching to a substrate
or in a heterostructure, its total polarization is the sum of its intrinsic spontaneous polarization,
$P_{sp}$ (that is the polarization that it would have in an unstrained bulk sample) plus the 
polarization induced as a result of the strain, $P_{pz}(\epsilon)$. 
The strain-induced component depends strongly on the strain, $\epsilon$, and hence on the lattice 
mismatch between the epitaxial layers; in the linear regime it is related to the piezoelectric 
tensor, $e$, by
\begin{equation}
P_{{pz}(i)} = \sum_{j}{e_{ij}}\epsilon_j .
\label{polar_eqn}
\end{equation}
Both the intrinsic and strain-induced contributions can be calculated 
accurately from first principles calculations\cite{Vander-pol1,Vander-pol2,Resta,Bernardini},
and have been shown to be large in the III-V nitrides \cite{Bernardini}.

We begin by calculating the polarization and piezoelectric constants ($e_{33}$ and $e_{31}$)
at the calculated LDA equilibrium structures for MgO ($a = 3.26$ \AA, $c/a = 1.51$, $u=0.398$), 
ZnO ($a = 3.20$ \AA, $c/a = 1.61$, $u=0.378$) and CdO ($a = 3.60$, $c/a = 1.55$, $u=0.391$); 
our results are reported in Table~\ref{eqb-pol}. For comparison we also report the literature 
values calculated for the nitrides at their theoretical LDA lattice parameters and $u$ values
\cite{Bernardini3}.
(Note that our reported polarization for LDA ZnO (-0.048 C/m$^2$) differs slightly from the literature 
value (-0.057 C/m$^2$) \cite{Bernardini} because we use our pseudo-SIC approach; if we repeat 
our calculations using the LDA we reproduce the earlier value. Likewise, our pseudo-SIC piezoelectric 
constants are slightly different from literature LDA values\cite{Bernardini,dalcorso}.)

\begin{table}
\begin{center}
\begin{tabular}{c|c c c c}
\hline
       &     &    $P$   & $e_{33}$ &  $e_{31}$  \\ \hline \hline
       & MgO &  -0.173  &    2.26  & -0.38      \\
II-VIs & ZnO &  -0.048  &    1.34  & -0.57      \\
       & CdO &  -0.099  &    1.67  & -0.48      \\ \hline
       & AlN &  -0.081  &    1.46  & -0.60      \\
III-Vs & GaN &  -0.029  &    0.73  & -0.49      \\
       & InN &  -0.032  &    0.97  & -0.57      \\ \hline
\end{tabular}
\caption {Polarizations, $P$, and piezoelectric constants, $e_{33}$ and $e_{31}$, in C/m$^2$, for II-VI
oxides (this work), and III-V nitrides (from  Ref. \onlinecite{Bernardini}), calculated at the theoretical 
LDA equilibrium lattice parameters.}
\label{eqb-pol}
\end{center}
\end{table}

\begin{figure}[htbp]
\begin{center}
\includegraphics*[width=0.45\textwidth]{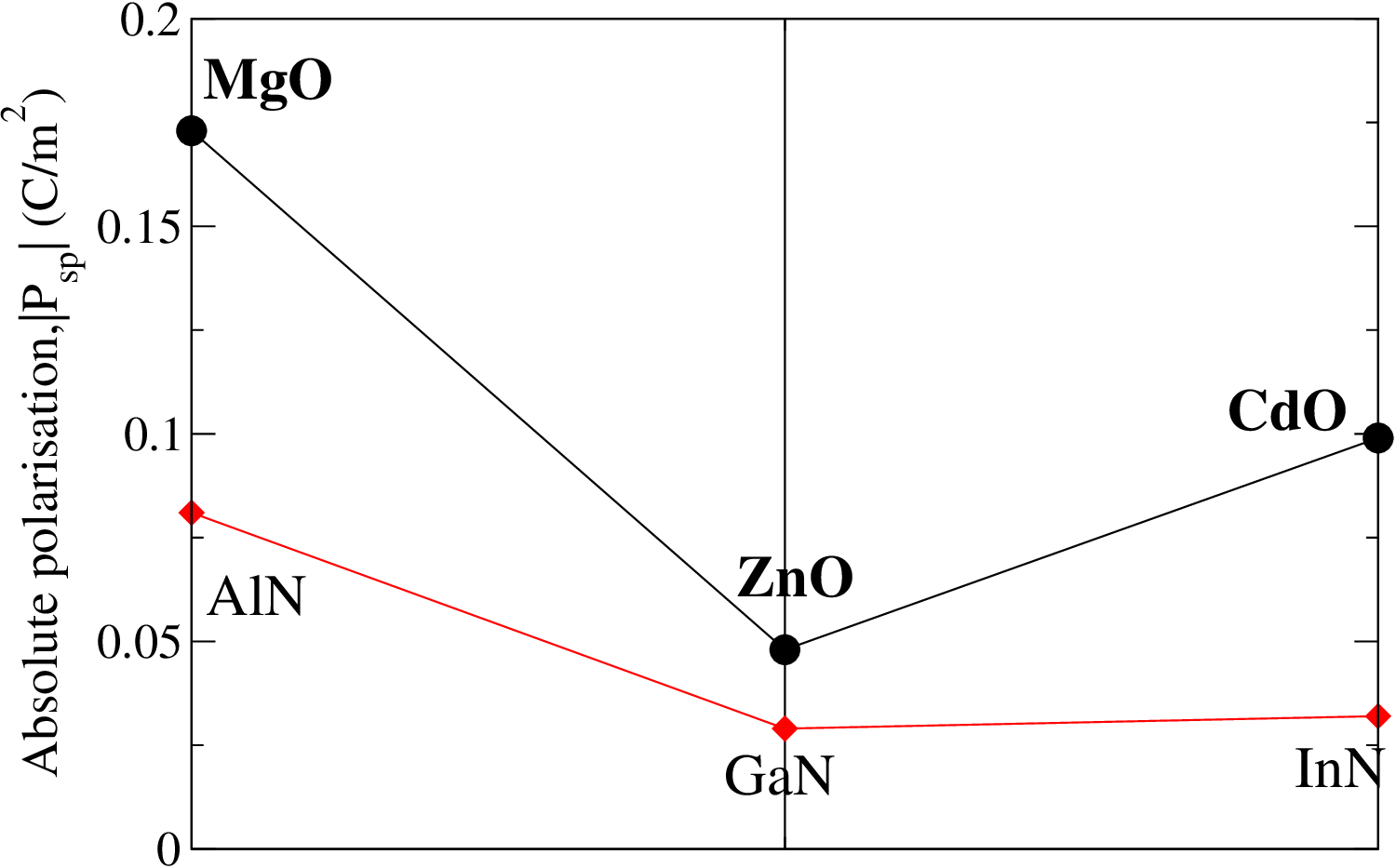}
\end{center}
\caption{Calculated absolute polarization values of the II-VI oxides ZnO, MgO and CdO (this work) and their 
III-V nitride analogues GaN, AlN and InN (from Ref. \onlinecite{Bernardini}).}
\label{alloy-pol}
\end{figure}

Notice that the polarizations are larger in every case for the oxide 
materials than for their nitride analogues. Perhaps more importantly, the polarization {\it gradients}
between the end-point materials are considerably larger in the ZnO-based system. This is illustrated in 
Fig.~\ref{alloy-pol}, where we compare our polarizations for MgO, ZnO and CdO with the literature values 
for AlN, GaN and InN, all at the theoretical equilibrium structures. The straight lines indicate the 
Vegard's Law projections for the polarizations of intermediate alloy compositions. 

In principle,
the polarization in any other strain state can then be obtained from the equilibrium values shown in
Table~\ref{eqb-pol} using Eqn.~\ref{polar_eqn}. However, since the experimental bond lengths 
depend sensitively on the details of the structure, and could in fact be distorted beyond the linear 
regime described by Eqn.~\ref{polar_eqn}, we also calculate the polarization from first principles in 
two other limits: 
First, at the experimental ZnO lattice parameters and atomic positions ($a=3.25$ \AA, $c/a=1.60$, $u=0.382$
\cite{LB}).  This represents the situation in which only small amounts of Mg or Cd are doped into ZnO 
so that the structure does not change significantly. And second, we constrain the in-plane lattice parameter 
\textbf{$a$} to that of experimental ZnO, and allow the structure to relax perpendicular to the plane (along
the $c$ direction) to simulate the situation in epitaxial films. 

Our results are shown in Table~\ref{zno-expt-pol}. It is clear from the first column (calculated using 
the ZnO experimental coordinates) that, when the ions are constrained to have the same ionic positions, 
the polarizations of the three materials are rather similar. Therefore differences in polarizations in 
real systems must result primarily from the relaxations of the ions. 
The second column of data in Table~\ref{zno-expt-pol} indicates that when MgO is allowed to relax out
of plane, there is a large increase (almost double) in the polarization; this is caused by a contraction 
of the $c$ parameter during the structural optimization ($c/a$ reduces from 1.60 to 1.52 and $u$ changes
from 0.382 to 0.395).
Similarly the the large (factor of six) decrease which we find for CdO is caused by an out-of-plane expansion 
($c/a$= 1.87, $u=0.357$) to accommodate the applied in-plane compression. 
However the polarization of ZnO is almost unchanged from that at the experimental coordinates, because
there is negligible change in the $c$ lattice parameter on relaxation.
Note also that the polarization of MgO is similar to that at its calculated equilibrium parameters
(Table~\ref{eqb-pol}), because the 
calculated equilibrium parameters are not dissimilar to those with the in-plane constraint. 

\begin{table}
\begin{center}
\begin{tabular}{ccc}
\hline
 &\multicolumn{2}{c}{$P$}   \\
\hline
    &   (i)  &  (ii)  \\
MgO & -0.067 &  -0.152 \\
\hline
ZnO & -0.073 & -0.081 \\
\hline
CdO &-0.073  & -0.013  \\
\hline
\end{tabular}
\caption {Polarization values for the II-VI oxides calculated (i) at the ZnO experimental atomic positions and
lattice parameters, and (ii) with the $a$ parameter constrained to the experimental ZnO value, and the theoretical
relaxed $c$ parameters and atomic positions.}
\label{zno-expt-pol}
\end{center}
\end{table}

Finally, (Table~\ref{elastic_constants}) we report the elastic constants (C$_{11}$, C$_{12}$, C$_{33}$, C$_{13}$, 
and C$_{44}$) obtained by calculating the change in energy with strain for five different strain configurations. 
The details of the strain configurations used and the procedure for extracting the elastic constants are described 
in Ref.  \onlinecite{Wright}. For comparison, the experimental values for ZnO, obtained using the Brillouin light
scattering technique on polycrystalline films \cite{Carlotti} are also included in the table. Note that the agreement
between our calculated values and the measured values for ZnO is very good. We expect our 
calculated values for hypothetical wurtzite-structure MgO and CdO to be similarly appropriate.

\begin{table}
\begin{center}
\begin{tabular}{l c r r r r r}
\hline
          &     & $C_{11}$ & $C_{12}$ & $C_{33}$ & $C_{13}$ & $C_{44}$  \\
\hline
          & MgO &   222    &    90    &   109    &    58    &   105     \\
\hline
calculated& ZnO &   217    &   117    &   225    &   121    &    50     \\
\hline
          & CdO &   150    &   108    &   105    &    61    &    47     \\
\hline \hline
measured  & ZnO &   206    &   118    &   211    &   118    &    44     \\
\hline
\end{tabular}
\caption {Elastic constants for wurtzite-structure MgO, ZnO and CdO calculated at
the theoretical equilibrium lattice parameters and atomic positions (this work) and 
measured using Brillouin light scattering (Ref. \onlinecite{Carlotti}).}
\label{elastic_constants}
\end{center}
\end{table}
 
In summary, we have calculated the spontaneous polarizations, piezoelectric coefficients and elastic
constants for wurtzite-structure MgO, ZnO and CdO. For ZnO our results agree well with previous
calculations and with experimental data; for MgO and CdO (which are experimentally inaccessible
in the wurtzite structure) we provide predictions. We find that the absolute polarizations and the 
polarization gradients between end-point compounds are larger in the ZnO system than in the 
GaN system.


\smallskip
This work was supported by the Department of Energy, grant number
DE-FG03-02ER45986, and made use of MRL central facilities, supported
by the the National Science Foundation under the
Award No. DMR00-80034. We thank Profs. Chris van de Walle and Jim Speck for
invaluable discussions.

\bibliography{ZnO-polar-final}

\end{document}